\documentclass[twocolumn,showpacs,preprintnumbers,amsmath,amssymb]{revtex4}

\usepackage{graphicx}
\usepackage{dcolumn}
\usepackage{bm}

\begin{document}

\title{From Space and Time to a Deeper Reality\\
as a Possible Way to Solve Global Problems}

\author{Galina Korotkikh}
\email{g.korotkikh@cqu.edu.au}

\affiliation{School of Information and Communication Technology\\
CQUniversity, Mackay \\ Queensland, 4740, Australia}

\author{Victor Korotkikh}

\email{v.korotkikh@cqu.edu.au}

\affiliation{School of Information and Communication Technology\\
CQUniversity, Mackay \\ Queensland, 4740, Australia}

\begin{abstract}

To deal with global problems we suggest to consider complex systems
not in space and time, but in a possible deeper reality, i.e., the
hierarchical network of prime integer relations. Encoded by
arithmetic through the self-organization processes the hierarchical
network appears as the mathematical structure of one harmonious and
interconnected whole. Remarkably, the holistic nature of the
hierarchical network allows to formulate a single universal
objective of a complex system defined in terms of the integration
principle. We propose that by the realization of the integration
principle the Earth system could be transformed to become an
integrated part of a larger system with more capacity and energy to
sustain life. Significantly, based on integers and controlled by
arithmetic only the hierarchical network has a unique potential to
provide an irreducible common ground fully trusted by different
parties and helping to reveal a higher collective purpose.

\end{abstract}


\maketitle

\section{Introduction}

Paradoxically, in the world experiencing rapid progress in science
and technology the understanding of growing problems is limited.
They become numerous and affect us all. For example, the recent
financial crisis has sharply revealed that the global financial
system shaped and entangled into a complex entity lacks the clarity.
Moreover, the longtime debates on the climate change have not
resulted in a comprehensive agreement. There is still no common
understanding of this complex problem, where experimental evidence
can convince different parties. Despite that, we can hope for a
better future.

An approach aiming to provide a solution may lead to an interesting
speculation. To be of practical interest it has to meet strong
requirements. In particular, the approach has to be able to explain
the observed phenomena and at least be based on an irreducible
foundation. Once the approach is in agreement with observation and
calls for a collective action to solve global problems, the proposed
action could be accepted and fully trusted by different parties. In
this context we suggest to consider complex systems not in space and
time, but in a possible deeper reality, i.e., the hierarchical
network of prime integer relations. Encoded by arithmetic through
the totality of the self-organization processes, the hierarchical
network appears as the mathematical structure of one harmonious and
interconnected whole. Remarkably, based on integers only the
hierarchical network can secure the foundation and picture complex
systems in irreducible terms \cite{Korotkikh_1}-\cite{Korotkikh_5}.

Moreover, the holistic nature of the hierarchical network allows to
formulate a single universal objective of a complex system defined
in terms of the integration principle
\cite{Korotkikh_6}-\cite{Korotkikh_8}. We suggest that by the
realization of the integration principle the Earth system could be
transformed to become an integrated part of a larger system with
more capacity and energy to sustain life. We also discuss how the
integration principle of the Earth system could be effectively
realized.

\section{The Hierarchical Network of Prime Integer
Relations as a Possible Deeper Reality}

To introduce the hierarchical network we present results based on
the description of complex systems in terms of self-organization
processes of prime integer relations
\cite{Korotkikh_1}-\cite{Korotkikh_5}.

Rather than in space and time, the description suggests to consider
complex systems on a new stage, i.e., the hierarchical network of
prime integer relations. The hierarchical network can be defined by
two equivalent forms, i.e., arithmetical and geometrical.
Remarkably, in the hierarchical network arithmetic and geometry are
unified together, yet play the different roles. For example, while
the arithmetical form sets the relationships between the parts of a
complex system, the geometrical form makes possible to measure their
effects on the parts.

In the arithmetical form the hierarchical network comes into
existence by the totality of the self-organization processes of
prime integer relations. Starting with the integers the processes
build the hierarchical network under the control of arithmetic as
one harmonious and interconnected whole, where not even a minor
change can be made to any of its elements.

In its turn, a prime integer relation of a level is also built by a
process first from integers and then from prime integer relations of
the levels below. Notably, a prime integer relation can be seen not
only as an indivisible whole but as a complex system itself. Indeed,
a prime integer relation is made of integers as the ultimate
building blocks with the relationships set by arithmetic, where each
and every element in its formation is necessary and sufficient for
the prime integer relation to exist. That is why we call such
integer relations prime.

In the hierarchical network a complex system is defined by a number
of global quantities, which remain invariant under certain
self-organization processes. As a result, a complex system can be
characterized by the hierarchical correlation structures determined
by the self-organization processes of prime integer relations. The
correlation structures operate through the relationships emerging in
the formation of the prime integer relations. Since a prime integer
relation expresses a law between the integers, the complex system
become governed by the laws of arithmetic realized through the
self-organization processes of prime integer relations.

A hierarchical structure of prime integer relations can be produced
by a system of Diophantine equations. When a hierarchical
correlation structure become operational, a corresponding solution
of the Diophantine equations simultaneously gives rise to the prime
integer relations.

In the geometrical form the processes a complex system is defined by
are isomorphically expressed  in terms of transformations of
two-dimensional patterns. As a result, the hierarchical correlation
structures of the complex system become represented by the
hierarchical structures of the geometrical patterns. Importantly,
this geometrizes the correlations as well as the laws of arithmetic
the complex system is determined and opens a way to characterize the
correlations and the laws of arithmetic by space and time as dynamic
variables. Therefore, the geometrical form allows to transform the
laws of a complex system in terms of arithmetic into the laws of the
system in terms of space and time.

In our description a complex system is nothing but a manifestation
of the underlying processes in the hierarchical network as well as a
transitory entity in the realization of the prime integer relations.
Since a prime integer relation can be fully represented by its
geometrical pattern, the elementary parts of a complex system come
into existence in the expression of their geometrical patterns.

The geometrical pattern of an elementary part can be defined by its
area and boundary curve. Once the boundary curve is specified by the
space and time variables of the elementary part and the area is
associated with the energy, the dynamics of the elementary part
become determined by its geometrical pattern. Therefore, an
elementary part may be seen as a boundary curve supplemented with a
corresponding number.

In the description the boundary curve of an elementary part at level
$l$ is given by a polynomial of degree at most $l$. The coefficients
of the polynomial, except the last one, can be seen as the quantum
numbers of the elementary part. Notably, the quantum numbers of the
elementary parts in a hierarchical correlation structure are all
conserved.

Surprisingly, in the hierarchical network of prime integer relations
a complex system works harmoniously well. In particular, once a
hierarchical correlation structure of a complex system become
operational, then through the corresponding prime integer relations
all parts become instantaneously connected and move to realize the
prime integer relations simultaneously. The effect of the
correlations may be different for the elementary parts, but for each
elementary part it is exactly as required to preserve the system.
Namely, the clocks of the elementary parts may tick differently, yet
precisely as determined by the prime integer relations.

Remarkably, the elementary parts of a correlation structure act as
the carriers of the laws of arithmetic with each elementary part
carrying its own quantum of a law fully determined by the
geometrical pattern. Therefore, the quantum of arithmetic law of an
elementary part defines the local spacetime and the energy of the
elementary part. This suggests an important perspective to use
elementary parts in the hierarchical network as fundamental entities
to construct different laws. For example, as the quanta of
arithmetic laws of elementary parts would be combined to take a form
in terms of the same number of space and time variables, a global
spacetime of the elementary parts could be produced.

Therefore, in the hierarchical network the laws of a complex system
are entirely given by laws of arithmetic and whatever form these
laws can take, they will still remain a manifestation of pure
arithmetic.

As a prime integer relation and thus its geometrical pattern can not
be changed even a bit, the hierarchical network may be viewed as
two-dimensional rigid bodies interconnected by their
transformations.

Significantly, all forces in the hierarchical network are managed by
the single "force" - arithmetic. It serves the special purpose to
hold the parts of a system together and possibly drive its formation
to make the system more complex. Therefore, in the hierarchical
network the forces do not exist separately, but through the
self-organization processes are all unified and controlled to work
coherently in the preservation and formation of complex systems.

These all explain our motivation to advocate the hierarchical
network as a possible deeper reality. Indeed, built on an
irreducible foundation the hierarchical network could be a reality
that demonstrates through the processes the power to create and
control complex systems. By recognizing this transformative power it
would be then important to understand whether this power can be
harnessed and used to solve global problems in particular.

\section{Integration Principle as the Master Equation of a Complex System}

Because the forces acting in space and time as well as their meaning
remain unknown, it is not clear how a single objective of a complex
system could be formulated there. By contrast, the holistic nature
of the hierarchical network allows to formulate a single universal
objective defined in terms of the integration principle of a complex
system \cite{Korotkikh_6}-\cite{Korotkikh_8}:
\medskip

{\it In the hierarchical network of prime integer relations a
complex system has to become an integrated part of the corresponding
processes or the larger complex system.}
\medskip

Significantly, the integration principle determines the general
objective of the optimization of a complex system in the
hierarchical network.

The geometrical form of the description plays a special role in the
realization of the integration principle. In particular, the
position of a system in the corresponding processes can be
associated with a certain two-dimensional shape, which the
geometrical pattern of the optimized system has to take to satisfy
the integration principle. Therefore, in the realization of the
integration principle it is important to compare the current
geometrical pattern of the system with the one required for the
system by the integration principle. Since the geometrical patterns
are two-dimensional entities, the difference between their areas can
be used to estimate the result.

Moreover, the fact that in the hierarchical network processes can
progress level by level in one and the same direction and, as a
result, make a system more and more complex, may suggest a possible
way for the realization of the integration principle. In particular,
as the complexity of a system increases level by level, the area of
its geometrical pattern may monotonically become larger and larger.
Consequently, with each next level $l < k$  the geometrical pattern
of the system would fit better into the geometrical pattern
specified by the integration principle at level $k$ and deviate more
after. In its turn, the performance of the optimized system could
also increase to attain the global maximum at level $l = k$.

Therefore, as the area of the geometrical pattern of a system and
its complexity would increase with each level $l$, the performance
of the system might behave as a concave function of the complexity
with the global maximum at level $k$ specified by the integration
principle.

Computational experiments have been conducted to test this
prediction. Remarkably, they support the claim and show that the
integration principle of a complex system could be efficiently
realized. Moreover, the experiments indicate that in the
hierarchical network NP-hard problems could be avoided
\cite{Korotkikh_6}-\cite{Korotkikh_8}.

According to the description the Universe we observe through the
ordinary senses  may be just a projection of certain
self-organization processes defining a complex system in the
hierarchical network \cite{Korotkikh_1}. This projection of three
dimensions of space and one dimension of time may be effective to
achieve certain objectives, yet not sufficient to solve global
problems. Therefore, we suggest to consider the Universe directly in
the hierarchical network with the Earth system \cite{Kleidon_1} as
its integrated part.

In the hierarchical network the integration principle can be seen as
the master equation of a complex system. It specifies the position
of a complex system in the corresponding processes and could
determine the forces acting on the system, its physical constants
and parameters. Consequently, we suggest to consider the solution of
global problems through the realization of the integration principle
of the Earth system. In particular, we propose that through the
integration of human activities and the Earth itself with the
Universe as a complex system in the hierarchical network global
problems could be resolved.

In the realization of the integration principle of the Earth system
two main questions arise: first, how to specify the Universe by
self-organization processes in the hierarchical network and second,
how to characterize the Earth system.

Remarkably, self-organization processes of prime integer relations
demonstrate properties of scale invariance. In particular, we have
discovered a process where all levels can be subdivided into the
groups of three successive levels, so that in a group the process
can be characterized by a series of approximations specified at the
lower groups of levels \cite{Korotkikh_1}.

Therefore, once scale invariant properties of the processes the
Earth system belongs to would be established, the characterization
of the Earth system could be effectively simplified to realize the
integration principle.

To this end, the information about the processes at the lowest group
of levels would be very important. To specify the processes at the
lowest group of levels we suggest to represent the standard model of
elementary particles \cite{Hoddeson_1}-\cite{Quigg_1} in terms of
the hierarchical network. For this purpose, the large symmetry of
the hierarchical network and data on the energies of its elementary
parts could be used.

As under the processes the Earth system could be involved in the
formation of a more complex system, it would be then important to
understand whether the transformation takes place and, if it does,
specify it. Significantly, according to the description, in the
formation of a complex system the energy may increase
\cite{Korotkikh_1}. Therefore, if the Universe experiences a further
formation, the description predicts that the new energy would be
generated and this could result in the acceleration of the Universe.

It is remarkable that the accelerated expansion of the Universe has
been recently observed in physical cosmology \cite{Riess_1},
\cite{Perlmutter_1}. The observational evidence may support the idea
of the formation and the transformation of the Earth system itself.
Notably, as the acceleration of the Universe is attributed to dark
energy \cite{Peebles_1}, the description has a potential to explain
this energy in terms of the processes and thus arithmetic.

Since the description could characterize the formation of the
Universe as well as the released energy quantitatively, the
experimental data on the acceleration of the Universe would be
useful to specify the transformation of the Earth system.

In order to realize the integration principle of the Earth system it
would be required to convert information about human activities in
terms of space and time into the information in terms of the
hierarchical network.

Usually, in space and time information about a complex system is
given by time series that can be used to characterize correlations
through the variance-covariance matrix. Based on computational
results \cite{Korotkikh_7}, \cite{Galina_1}, \cite{Galina_2} we
suggest to apply eigenvalues and eigenvectors of variance-covariance
matrices to obtain information about a complex system in the
hierarchical network from the information about the system in space
and time. In particular, this could specify the correlation
structure and the geometrical pattern of the Earth system by using
corresponding time series.

In this regard our method resonates with the idea in econophysics to
characterize a financial system by eigenvalues and eigenvectors of
variance-covariance matrices \cite{Mantenga_1}-\cite{Plerou_1}. The
idea has been motivated by the success of random matrix theory in
explaining the statistics of energy levels of complex quantum
systems \cite{Mehta_1}, \cite{Dyson_1}. Although random matrix
theory has opened a way to study the structure of a complex system,
yet so far it lacks a general framework to define the structure of
all possible structures and, more importantly, its geometry. By
contrast, in our description the structure of a complex system is an
integrated part of the hierarchical network. It is defined by the
correlation structures determined by self-organization processes of
prime integer relations, while its geometry comes from the
corresponding two-dimensional patterns.

\section{Conclusions}

In the paper we have suggested to consider global problems from the
perspective of the hierarchical network of prime integer relations
as a possible deeper reality and seek their solution in terms of the
integration principle of the Earth system.

The practical realization of the integration principle of the Earth
system would require the development of a Global Integrating System
with two basic components, i.e., the navigation and control systems.

The purpose of the navigation system would be to specify the
processes the Earth system belongs to and provide information for
the realization of the integration principle. In general, the
navigation system could be based on a network of existing particle
accelerators, space telescopes and satellites. In particular, they
could be used to identify the processes at the boundary levels.

The idea of the control system would be to ensure that human
activities are consistent with the integration principle of the
Earth system. For this purpose the control system should be designed
to convert the information about human activities in terms of space
and time into the information in terms of the hierarchical network,
then compare the result with the output from the navigation system
and make the correction if needed.

The control system could emerge through the development of the
global financial system as well as cyber infrastructures of global
supply chains and economic production
\cite{Korotkikh_9}-\cite{Korotkikh_11}.

The development of the Global Integrating System would require a new
type of international cooperation. Realistically, the cooperation is
unlikely unless it is based on an irreducible ground fully trusted
by different parties and helping to reveal a higher collective
purpose. Remarkably, based on integers and controlled by arithmetic
only the hierarchical network of prime integer relations has a
unique potential to provide such a common ground.

The practical realization of the integration principle of the Earth
system might be viewed as the first attempt to extend human
scientific inquiry and experimental activity into the new reality,
i.e, the reality, where parts of a complex system could be
instantaneously connected irrespective of the distances they may be
far apart, the reality, where a complex system could be efficiently
managed as one whole with its pasts, presents and futures all united
at once.

Significantly, the reality of the hierarchical network of prime
integer relations promises new sources of energy. In the
hierarchical network it is arithmetic that controls energy to be
conserved or generated in the exact amount. This source may be
associated with dark energy, yet it would be a completely different
story to be able to use it with all technological consequences.

Likewise, the hierarchical network could be seen as a source of laws
to achieve different objectives. For a given objective the
hierarchical network could be used to generate self-organization
processes with relevant laws of arithmetic to be processed into the
required form by constructing a corresponding global spacetime.

Finally, we speculate about the existence of a Super Intelligence
that thinks well to produce the integers and thus the hierarchical
network of prime integer relations, where processes can be
comprehensible and efficiently controlled. As soon as the
transformative power of the hierarchical network would be
successfully tested, humans could become an integrated part of the
Super Intelligence consciously operating in the reality.

This promises humans a special role in the reality one day they may
call their home.

\bibliography{apssamp}

\end{document}